\def\simlt{\lower.5ex\hbox{$\; \buildrel < \over \sim \;$}}
\def\simgt{\lower.5ex\hbox{$\; \buildrel > \over \sim \;$}}
\def\simpropto{\lower.2ex\hbox{$\; \buildrel \propto \over \sim \;$}}

\documentclass{iaus}
\usepackage{graphicx}

\title[Formation and evolution of disk galaxies
] 
{Formation and evolution of disk galaxies}

\author[Joseph Silk]   
{Joseph Silk}

\affiliation{Dept. of  Physics, Oxford Univ.,  Denys Wilkinson Bldg., 1 Keble Rd., Oxford OX1 3RH UK \\email: {\tt silk@astro.ox.ac.uk}}

\pubyear{2008}
\volume{254}  
\pagerange{119--126}
\jname{The Galaxy Disk in Cosmological Context}
\editors{J. Andersen,
J. Bland-Hawthorn \&
B. Nordstršm, eds.}

\def\simlt{\lower.5ex\hbox{$\; \buildrel < \over \sim \;$}}
\def\simgt{\lower.5ex\hbox{$\; \buildrel > \over \sim \;$}}
\def\simpropto{\lower.2ex\hbox{$\; \buildrel \propto \over \sim \;$}}

\begin{document}

\maketitle

\begin{abstract}
Global star formation is the key to understanding galaxy disk formation. This in turn depends on gravitational instability of  disks and continuing gas accretion as well as minor merging. A key component is feedback from supernovae. Primary observational constraints on disk galaxy formation and evolution include the Schmidt-Kennicutt law, the Tully-Fisher relation and the galaxy luminosity function. I will review how theory confronts phenomenology, and discuss future prospects for refining our understanding of disk formation.
\end{abstract}

\firstsection 
\section{Introduction}
There are three interrelated empirical scaling functions 
that relate all disk galaxies and are fundamental to understanding their formation. These are
the Tully-Fisher law, 
the Schmidt-Kennicutt law, 
and the galaxy luminosity function. 
If we can ever understand these, we will have come a long way towards understanding disk galaxy formation.  I will review some of the issues that arise when theory confronts phenomenology. 
Star formation and mass assembly are the primary issues to be understood. Although intimately connected with star formation, I will mostly avoid chemical evolution constraints in this discussion, in part because of their strong sensitivity to adoption of an IMF and because they are covered elsewhere in this conference. 
\section{Tully-Fisher relation}

The Tully-Fisher relation  is a long known correlation between disk luminosity and maximum rotational velocity.
The principal current issues in understanding the Tully-Fisher (TF) relation involve angular momentum loss 
and adiabatic contraction. 
The I-band Tully-Fisher relation, preferred to the B-band data because of the lower scatter,
 provides a challenge for understanding current epoch disks.
The contribution of turbulence and rotation to line widths reduces TF scatter and is
measured to $z\sim 1.$ No significant evolution in the TF relation is seen (Kassin et al. 2007).
Even galaxies as far back as $z\sim 2$ (Bouch\'e et al. 2007) remain on the  TF law as measured locally.

Loss of angular momentum of the baryons to the dark halo was discovered in numerical simulations
(Steinmetz \& Navarro 1999).
Strong SN feedback can reduce angular momentum loss (Governato et al. 2007). The observed normalisation  can be obtained in the inner disk. This success comes at a price, however. One concern is that of the  implications for the galaxy luminosity function at low luminosities. Another is disk thickness. 
The simulated disks are thicker than observed disks  although the numerical resolution is inadequate for a real confrontation with data. 

The radial structure is a more pressing issue.
Lack of evolution is observed in disk sizes  at $z<1$,
in contrast to the remarkable  compactness found at  $z\sim 2-3$ (Buitrago et al. 2008).
Adiabatic contraction of the dark halo due to the self-gravity of the collapsing baryons  helps account for this by modifying the effective size evolution at early epochs (Somerville et al. 2008). 
However the response of the dark halo
to baryon dissipation  by adiabatic contraction means that once the rotation curves are  properly corrected, the asymptotic rotation velocity fails to match the observed TF zero point (Dutton et al 08).
An equivalent symptom of the TF issue is that the predicted mass-to-light ratio is too high.

\section{The Schmidt-Kennicutt law}

The disk star formation rate is coupled to the gas surface density by the empirical relation
$$SFR=0.02 (GAS \ \ SURFACE \ \  DENSITY)/t_{dyn}.$$
Remarkably, this relation 
fits quiescent as well as  star-bursting galaxies. The individual star-forming complexes in M51 also lie along this line (Kennicutt et al. 2008). This is more of a mystery and suggests that local processes control the efficiency of star formation.  

We may regard the Schidt-Kennicutt (SK) law as specifying the disk mode of star formation. It is
motivated globally by the gravitational instability of cold disks.
Feedback from supernovae is crucial to avoid too high a star formation rate. Momentum balance yields 
$$SFE   =   \sigma _ {gas}  v_{cool}  m^*_{SN} / E^{SN}_{ initial}\approx 0.02.$$ The global injection of momentum is balanced by local dissipation. Here $m^*_{SN} $ is the mass in forming stars required to produce a supernova of Type II, approximately 150 M$_\odot $ 
for a Chabrier IMF, and $E^{SN}_{ initial}\approx 10^{51}$ ergs is the initial kinetic energy in the supernova explosion.

In this way, a near steady state of cloud motions is obtained at the disk interstellar  cloud velocity dispersion of $\sigma\approx 10\rm km/s.$ This corresponds to the peak in the diffuse ISM cooling curve. The local control of momentum input and dissipation is to be found in exploring cloud disruption once massive stars form, develop HII regions  and explode.
The explanation is derived from the local physics of cloud efficiency which controls the efficiency of  star formation, that is, the fraction of gas converted into stars within a specified time-scale. The relevant local time-scale is molecular cloud lifetime.  The combination [(SFR $\times$ dynamical  time) /gas density  is approximately  constant as a function of  gas density, and amounts to about 0.02 (Krumholz \& Tan 2007).

There are several unexpected developments for the Schmidt-Kennicutt (SK)  law.
Star-forming complexes in M51 individually lie on the SK law, albeit with increased dispersion and a slight steepening of the slope (Kennicutt et al. 2007).
Extreme starbursts fit  the SK law both in the nearby universe (Sajina et al. 2006) and to $z\sim 2$ for  SMGs
(Bouch\'e et al  2007). The SK law is dominated by the molecular content of the galaxies. This is not too surprising as stars form in molecular clouds. The 
total molecular mass measured via CO amounts to $\sim$10 \% of the HI in LMC, M33. In the MWG, the atomic and molecular gas masses are comparable, and many ultraluminous infrared-emitting galaxies are molecule-dominated
(Gao \& Solomon 2004). Use of molecular indicators that trace the densest gas such as HCN reveal a linear law whereas use of CO, which traces the least dense molecular gas, prefers the SK law
(Komugi et al. 2005). The change of slope is most simply understood in terms of the excitation thresholds 
(Krumholz \& Thompson 2007), with the dominant mass in molecular gas, traced by CO, representing 
the physics of the self-gravitating cold disk instabilities and cloud collisions that drive the SK law. 
 
A more quantitative model of ISM turbulence comprises cloud collisions, with stirring by gravitational instabilities and by SNe momentum injection. 
Let a typical cloud have pressure $p_{cl}$ and surface density
$\Sigma_{cl}.$ We expect star-forming clouds to be marginally
self-gravitating, so that $p_{cl}=\pi G\Sigma_{cl}^2.$ They are
also marginally confined by ambient gas pressure, $p_g=\rho_g\sigma_g^2.$ 
Clouds certainly form this way, and this state  may be maintained if the cloud covering factor 
is of order unity. This guarantees  that cloud collisions occur on a local dynamical time-scale.
We have 
\begin{eqnarray}
p_g = \pi G \Sigma_g \Sigma_{tot} = p_{cl} = \pi G {\Sigma_{cl}}^2 , \end{eqnarray}
from which
$\Sigma_{cl} = (\Sigma_{tot} \Sigma_g)^{1/2}.$
Then the covering factor $\Omega_{cl}$ of clouds in the disk is directly inferred to be 
$ \Omega_{cl}= {\Sigma_g / \Sigma_{cl}} = {\left(\Sigma_g / \Sigma_{tot} \right)}^{1/2}.$
The cloud
collision time-scale is
$t_{coll}=({\Sigma_{cl}H})/({\Sigma_{g}\sigma_g}),$ where the scale
height $H^{-1}=({\pi G\Sigma_{tot}})/{\sigma_g^2}$,  
and $\sigma_g$ is the cloud velocity
dispersion.
The collision time can also be expressed as $t_{coll} = {\Omega_{cl}}^{-1} t_{cross}$
with $t_{cross} = {H/{\sigma_g}},$
which becomes
$ t_{coll} ={\left( {\Sigma_{tot} / \Sigma_g}\right)^{1/2} \left(H / \sigma_g\right)} .$

Assume the disk star formation rate is self-regulated by supernova feedback which drives the cloud velocity dispersion. Momentum balance gives
\begin{eqnarray}\dot\Sigma_\ast({E_{SN}}/({m_{SN}^\ast v_c}))=
f_c{\Sigma_g\sigma_g}/{t_{coll}}.\end{eqnarray}
Here $f_c$ is the cloud volume filling factor, which can be expressed in terms of porosity $Q$ as 
$f_c=e^{-Q}.$ Also
$v_c$ is the velocity at the onset of strong cooling. Canonical numbers used throughout are $m_{SN} ^\ast= 150 \rm M_{\odot}$  and $v_c= 400\, \rm kms^{-1}$.
 
We can rewrite this expression as 
\begin{eqnarray}
\dot\rho_\ast=\epsilon_{SN}f_c \sqrt{G\rho_g}\rho_g 
\end{eqnarray}
with $\epsilon_{SN}=({m_{SN}^\ast v_c\sigma_g}){E_{SN}}^{-1}({p_g}/{p_{cl}})^{1/2}.$ 
 For disks, we obtain 
\begin{eqnarray*}
{\dot{\Sigma_*}} = f_c[\left( \pi G \right) \left(\Sigma_{tot}\right)^{1/2} (({m_{SN} v_c )/ {E_{SN} }})][{p_g / p_{cl}}]^{1 \over 2} {\Sigma_g}^{3/2} \\
= \epsilon_{SN}f_c\sqrt{f_g }
(R/H)
^{1 \over 2}\Sigma_{gas} \Omega,
\end{eqnarray*}
where
$\Sigma_{tot} =\Sigma_{g} +\Sigma_\ast.$
Here the disk gas fraction is $f_g \sim 0.1$,  and we use the disk radius-to-scale-height relation
$H/R=(\sigma_g/v_r)^2$ for a disk rotating at $v_r$ with $\Omega^2=G\Sigma_{tot}/R.$
Remarkably, although 
the preceding formula ignores  the multi-phase nature of the interstellar medium and the possibility of gas outflows (see below), one nevertheless manages to fit the 
Schmidt-Kennicutt relation (\cite[Silk \& Norman 2008]{SilN08}). Mergers, which  certainly drive starbursts,  will increase the cloud velocity dispersion. However there is self-regulation  between 
star formation efficiency $\epsilon_{SN}\propto\sigma$ and  and disk scale-height $H\propto\sigma^2$.
This accounts for the robustness of the SK law fit.

There is one important addition to the above discussion. The disk heats up dynamically unless a supply of cold gas is provided. Cold gas accretion is needed in order to maintain global disk instability and 
continuing star formation. Major mergers cannot provide the gas source, as they would be too destructive for thin disks.  Minor mergers provide a motivated source. Observations provide strong justification for an extensive reservoir of neutral gas, as seen in the 
M33 group (Grossi et al 2008). Many nearby  spirals  are found to be embedded within extensive envelopes of HI. In some cases,  such as that of  NGC 6946, there is direct evidence of angular momentum loss as the HI accretes into the star-forming inner galaxy
as well as of extensive fountain activity (Boomsma  et al. 2008).
In all cases, the 
low star formation efficiency is plausibly due to SNe feedback, although this may not be the only mechanism at work. For example, magnetic pressure may also play a role in globally controlling gas flows via the Parker instability. 

\subsection{The rate of star formation }
In reality,  we find a  multiphase interstellar medium which consists of  hot phase
($\sim 10^6$ K gas) as well as  atomic and molecular gas. The heating and cooling of this gas, as well as mass transfer between the different phases,  controls the rate of star formation.
A  simple porosity description of supernova feedback in a multiphase ISM (Silk 2001) provides an expression for the star formation rate in which porosity-driven turbulence is the controlling factor: 
$\dot\rho_\ast=Q m_{SN}({4\pi/ 3}R_a^3t_a)^{-1}.$
The shell evolution  is generally described by
(Cioffi, Bertschinger and McKee 1988)
$t=t_0E_{51}^{3/14}n_g^{-4/7}(v_c/v)^{10/7} $
and
$R=R_0E_{51}^{2/7}n_g^{-3/7}(t/t_0)^{3/10} ,$
where
$
v_0=413 \rm km/s,  \   R_0= 14\rm pc  \ {\rm and} \   t_0=1.3\times 10^4 \rm yr .$
Here $v_c=413 E_{51}^{1/8}n_g^{1/4}\lambda^{3/8} \rm km\,s^{-1}$ where the  cooling time-scale within a SN-driven shell moving at velocity $v_c$ is 
$t_c=v_c/\lambda\rho,$  $\lambda^{-1}=3m_p^{3/2}k^{1/2}T^{1/2}/\Lambda_{ff}$
and $\Lambda_{eff}(T)$ is the effective cooling rate ($\simpropto t^{-1/2}$ over the relevant temperature range).

The SNR expansion is limited by the ambient  turbulent pressure to be $\rho_g\sigma_g^2$, and we
identify  the ambient turbulent velocity dispersion $v_a$ with $\sigma_g.$
We obtain
$$\dot\rho_\ast=Qc_0^{-1}m_{SN}^\ast n_g^{13/7}E_{51}^{-15/14}(\sigma_g/v_0)^{19/7}$$
where
$c_0={4\pi\over 3}R_0^3t_0.$ 
Rearranging, we have 
\begin{equation}
\dot\rho_\ast=Q\sqrt{G\rho_g}\rho_g({\sigma_g}/{\sigma_{fid}})^{19/7},
\end{equation} where
\begin{eqnarray*}
\sigma_{fid}=({c_0G^{1/2}m_p^{3/2}v_0^{19/7}E_{51}^{62/49}}
{{m_{SN}^\ast }^{-1}
)^{7/19} n_g^{-1/14}}
\\
\approx 20 n_g^{-1/14} m_{SN,100}^{-0.37}E_{51}^{0.47}\rm km/s,
\end{eqnarray*}
where $m_{SN}^\ast $  is normalized to 100 M$_\odot $ and $E_{SN}$ to $10^{51}$ ergs.
We rewrite the star formation rate  as 
$\dot M_\ast=\epsilon_Q M_g/t_d,  $  where 
$\epsilon_Q=
Q({\sigma_g}/{\sigma_{fid}})^{2.7}f_g^{1/2}.$ Now feedback physics compels to identify 
$\epsilon_Q$ with $\epsilon_{SN}.$ It follows that   $Q\propto\sigma^{-1.7}.$ Hence porosity decreases with increasing turbulence.

Simulations of star formation  with SN feedback in a multiphase ISM confirm that porosity provides a good description of feedback.  The analytical formula fits the numerical simulations performed at high enough resolution to follow the motions of OB stars prior to explosion (Slyz et al. 2006). 
The star formation rate prescription has been tested in numerical simulations of a merger.
In the case of the Mice (NGC 4676 a,b), a pure density law fails to account for the extended nature of star formation. A turbulence prescription for the SK law gives a better fit to the observed distribution of young stars (Barnes 2004).

\subsection{Application to starbursts}
Starbursts are characterised by high turbulent velocities and strong concentrations of molecular gas.
Locally, the gas density where stars form  is characterised by molecular clouds. It is presumably the cloud concentration that is enhanced by the dynamics of tidal interactions, which includes both stirring by bars and mergers.
Now the porosity ansatz, if porosity self-regulates, leads to an expression for the rate of star formation:
$$SFR = POROSITY 
\times (TURBULENT \ \ PRESSURE)^{1.36}.$$
We have already noted that if 
turbulent velocities are  high, the  porosity must be low. This conspiracy, together with that between
gas disk scale-height and star formation efficiency parameter, helps explain why even extreme 
merger-induced starbursts stay on the Schmidt-Kennicutt law.

It turns out that the  molecular gas fraction is regulated by the turbulent pressure
(Blitz and Rosolowsky 2006).The molecular fraction is found to be approximately proportional 
to the pressure,  so that the star formation rate empirically is 
 $$\Sigma_{SFR} = 0.1 \epsilon\Sigma_g\left(p_{mol}\over p_0\right)^{0.92}\rm M_\odot pc^{-2}Gyr^{-1}.$$ Here $\epsilon$ is an empirical SK law fit parameter. The molecular fraction is $(p_{mol}/ p_0)^{0.92\pm 0.10}$, where  $p_0/k=4.3 \pm 0.6\times  10^4\rm cm^3 K.$
 Since $\Sigma_g\simpropto p_{mol}^{0.5},$ the inferred star formation rate  dependence on pressure is similar to that obtained at constant porosity.
 Disk simulations that include a prescription for star formation in molecular clouds 
 (Robertson and Kravtsov   2008) can reproduce the slope and dispersion of the SK law..

It is plausible that porosity should self-regulate. Low  $Q$ is associated with
enhanced turbulent pressure. The supernova remnant expansion  is  blocked, further enhancing the  pressure and squeezing the clouds. This triggers more star formation and supernovae  that result in   blow-out of fountains from the disk. Blow-out drives high $Q$.
If $Q$ is high, we obtain lower pressure  and infall is allowed. The 
infall in turn reduces $Q$, and drives further accretion that eventually results in star formation  and supernova remnants. This in turn enhances $Q$.

Another reason that $Q$ self-regulates is that it is insensitive to the strength of the turbulence.
The volume filling fraction of cold gas depends  only logarithmically on the 
Mach number (Wada and Norman 2007).

The resulting star formation is not expected to be monotonic in a porosity-driven prescription.
The time-scale for variation is of order the life-time of a molecular cloud complex to disruption by 
SN feedback, of order $10^7-10^8$ yr. A local burst-like behaviour is indeed seen in the solar vicinity
(Rocha-Pinto et al. 2000). The inner star-forming  regions of nearby galaxies also reveal non-monotonic behaviour. For example,  M100 has undergone a sequence of  ÒburstsÓ (Allard et al.  2006).

Supernova-driven outflows lead to  entrainment and loading of the cold ISM into the fountain and/or wind.  
The degree of loading depends on the ability to resolve Kelvin-Helmholtz instabilities. This is completely inadequate even in state-of-the-art simulations. 
Empirically, the load factor is a few. 
as measured in  NGC 1569 (Martin et al. 2002).

The fate of the outflowing gas is uncertain. The outflow rate is proportional to $Q$ times the star formation rate times the load factor. Since $Q$ is suppressed as $\sigma^{-1.7}$, it follows that 
outflows are suppressed in massive galaxies. One still expect fountains to recirculate gas from disk to halo and back to disk, as the halo gas cools.

Star formation is observed to continue below the Toomre threshold for disk gravitational  instabilities.
This is naturally explained in the cloud collision model where even the outer regions are dynamically populated with orbiting disk clouds. The star formation rate is most likely modulated by the molecular gas fraction which itself is controlled by the local UV
radiation field (Schaye 2004).
\subsection{
Some current issues in understanding the Schmidt-Kennicutt law}

We still need to attain a deeper understanding of the universality of the SK law. 
It prevails both locally as well as globally, and at high as well as at low redshift. It applies in regions of extremely high star formation, such as extreme starbursts, and in regions of very low average star formation rate, albeit somewhat suppressed, such as the outer regions of disks and DLAs.

\section{Galaxy luminosity function}

The final diagnostic, and relic, of disk formation that I will discuss is the galaxy luminosity function.
The case for outflows has been made in the context of the cold dark model, as a means of suppressing the numbers of dwarfs (Dekel and Silk 1986) and  for driving early chemical enrichment of the intergalactic medium as inferred from the LBG "missing metals"  problem (Pettini et al 1999) and 
wind-driven enrichment  simulations (Cen \& Ostriker 1999, Oppenheimer \& Dave 2006).
Early numerical simulations of individual winds (Mac Low \& Ferrara  1999) argued for incomplete mixing of supernova ejecta, and this problem has not yet been resolved via multiphase simulations.
Suppression of winds in massive galaxy potential wells was demonstrated by 
Springel and Hernquist (2003). Ejection of baryons  may be accomplished either in the assembly stage 
via dwarf winds, as is relevant here, or from massive spheroids via AGN-triggered outflows. The latter is a separate topic that itself merits an entire review.

An alternative, and complementary, means of dwarf suppression appeals to reionisation inducing mass loss from low mass dwarfs, with escape velocity $\simlt 20$ km/s. This has been implemented in simulations  but fails to account for the shape of the  observed Milky Way luminosity function of dwarf satellites within 280 kpc (Koposov et al. 2007). The observed effective power-law index is $\alpha=0.25$ from $M_V= -2.5$  to  $-18. $ Either there is a poor fit to the observed luminosity function (Somerville 2002), or the predicted survivors have far too high a central surface brightness (Benson et al. 2002). Tidal stripping cannot account for the discrepancy (Penarrubia,  Navarro \& McConnachie 2008).

There is an equally serious problem for giant galaxies. The problem here is that accretion of gas continues over a Hubble time, and the galaxies continue to grow. The net result is too many and too blue massive galaxies. Supernova feedback cannot resolve these problems, but AGN/quasar feedback is considered to provide the needed panacea by heating the infalling gas and quenching star formation.The relevant feedback is that associated with supermassive black hole growth, and is destined to yield an explanation of the Magorrian relation between spheroid velocity dispersion and black hole mass.

This is not the only problem for massive galaxies and their halos: there are others. For example, cooling flows occur in galaxy clusters at a rate that exceeds the observed rate of cool gas deposition. 
Additional entropy input,  most likely in the form of preheating,  is needed to reduce their role, in order  to avoid an excessive rate of star formation in the brightest cluster galaxies.  AGN are again the most likely culprit.
Late feeding of AGN and jet-driven outflows provide a non-localised heat source that complements the 
early role of quasars in spheroid formation..

CDM-motivated theory   does account for at least one observational result, namely the characteristic galactic mass of the Schechter function fit, 
$L_\ast \sim 3\times 10^{10}\rm L_\odot.$ This is based on the requirement that cooling within a dynamical time is a necessary condition for efficient star formation. The inferred mass-to-light ratio is reasonable, although one is left with the need to hide a significant fraction of the baryons.These are 
probably in the form of intracluster gas (Gonzalez, Zaritsky \& Zabludoff 2007), within the observational errors, albeit 
some may be hidden at large radii.

Modelling has succeeded in giving an approximate fit to the galaxy luminosity function.
A relatively significant amount of dust is needed  for the massive galaxies in addition to both quasar and AGN feedback, as well as reionisation and SNe feedback for the low mass galaxies
(Bower et al. 2006; Croton et al. 2006; De Lucia \& Blaizot 2007). The models are tuned to fit optical data.  The fitting process is challenged as new, improved data becomes available. For example, the 
recent NIR data (Smith et al. 2008) reveals how fragile semi-analytical modelling   has become. The UKIDSS  SEDs yield 
improved stellar mass estimates that are not particularly well fit by published models, for either low mass or massive galaxies.

\section{SEEKING A UNIFIED THEORY}

Feedback is essential for slowing down or even quenching star formation. One can make the case that feedback, in some circumstances, accelerates star formation. One of the most worrying issues is that of the IMF. Comparison of the UV rest frame-measured star formation history and the NIR-measured stellar mass assembly histories agrees at low redshift but diverges towards high redshift (Wilkins et al. 2008). The explanation is unlikely to be due to uncertainties in the dust extinction law at  high redshift, where dust corrections are relatively small, A variation in the IMF is one possible explanation, the IMF becoming progressively top-heavy towards high redshift. 

There is a long history of top-heavy IMF discussions, largely centered around the G dwarf problem. This explanation is now largely discredited, in favour of a metal-free gas accretion history. The chemical abundance ratio signatures require that the Milky Way IMF was 'nearly invariant with  time, place  and metallicity (Gilmore \& Wyse 2004). 
However more recently, similar explanations invoking a top-heavy IMF have been advanced to explain the excess in the submillimetre galaxy  counts (Lacey  et al.  2008),
the enrichment of the intracluster gas (Nagashima et al. 2005),
the failure of semi-analytical modelling to reproduce the star formation efficiency enhancement towards high redshift  (Dav\'e 2008),  the extragalactic light background (Fardal  et al. 2007), and 
the luminosity and colour evolution of massive cluster galaxies (van Dokkum 2008). All of these 
interpretations, some of which have been given esoteric names by their proponents such as 'bottom-light' or 'paunchy',  involve a partial suppression of the mass-carrying stars, below a solar mass, relative  to those responsible for the bulk of the light and/or chemical yields. 

The best direct evidence for a  truncated or top-heavy IMF was 
in the Arches star cluster (Stolte et al. 2005) 
but this disappeared with deeper data (Kim et al. 2007).  Only within the central 0.1pc of our galaxy, within the gravitational influence of the central supermassive black hole, is there a confirmed deficit of 
low mass stars relative to the observed massive stars (Figer 2008). This does perhaps support  arguments for a top-heavy IMF in extreme situations, such as in the broad emission line regions of quasars where supersolar metallicities are observed and inferred to have been generated at $z\simgt 8$ 
(Mathias and Hamann 2008). However the gas masses involved,  $\simlt 10^4\rm M_\odot$
(Baldwin et al. 2003), seem to 
 fall well short of what is needed if gas enriched by a top-heavy IMF were to have a  global impact on galaxy evolution. 

Rather than add new parameters to the semi-analytical box of tricks, the future is more likely to bring greatly refined numerical treatments. It is clear that we are woefully resolution-limited when it comes to
the 3-dimensional mixing of multiphase media that involves  energy and momentum injection in one or more of the gas phases. Our understanding of feedback is likely to change dramatically once our simulations match our aspirations. Consider the track-record: our insights into star formation, galaxy harassment and  ram pressure stripping all underwent near 180 degree reversals once adequate adaptive grid power was brought to bear. I suspect that we are barely scratching the surface  when it comes to understanding angular momentum transfer and AGN or even SNe feedback in a multiphase interstellar medium embedded within a protogalaxy.
\acknowledgements
I thank my collaborator Colin Norman for many  discussions which have inspired 
much of the work described here.
	


\end{document}